\documentclass[a4paper,11pt]{article}
\usepackage{pos}

\newcommand{\cV}{\mathcal{V}}
\newcommand{\cM}{\mathcal{M}}
\newcommand{\cD}{\mathcal{D}}
\newcommand{\bzero}{\textbf{0}}
\newcommand{\id}{\mathds{1}}

\providecommand{\cM}{\mathcal{M}}
\providecommand{\cV}{\mathcal{V}}
\providecommand{\om}{\overline{m\hspace{-.2ex}}\hspace{.2ex}}
\providecommand{\oM}{\overline{\!M}\mkern-2mu}

\providecommand{\oK}{\overline{\!K}\,}

\usepackage{easybmat}
\usepackage{multirow,bigdelim}

\usepackage[utf8]{inputenc}
\usepackage{amsmath}
\usepackage{amsfonts}
\usepackage{booktabs}
\usepackage{makecell}
\usepackage{multirow}
\usepackage{pdflscape}
\usepackage{textcomp}
\usepackage{gensymb}
\usepackage{bigstrut}
\usepackage{array}
\usepackage{enumerate}

\usepackage{graphicx}
\usepackage{hyperref}
\usepackage{mathbbol}

\usepackage{appendix}
\usepackage{verbatim}
\usepackage{makecell}

\usepackage{listings}
\usepackage{mathtools}
\usepackage{dsfont}
\usepackage{arydshln}
\usepackage{mathrsfs}

\usepackage{graphicx} 
\usepackage{booktabs} 
\usepackage{subfigure}
\usepackage{ae}
\usepackage{mathrsfs}
\usepackage{graphicx}
\usepackage{bbm}
\usepackage{latexsym}
\usepackage{dsfont}
\usepackage{amsfonts}
\usepackage{amsmath}
\usepackage[super]{nth}
\usepackage{cancel}
\usepackage{stackengine}
\usepackage{epsfig}
\usepackage{subfigure}
\usepackage{graphicx}
\usepackage{amsmath}
\usepackage{verbatim}
\usepackage{enumitem}

\title{Non-Unitary Mixing Matrices in Neutrino and Vector-like Quark Models}

\author*[a]{Pedro M. F. Pereira}

\affiliation[a]{Centro de Física Teórica de Partículas – CFTP\\
Instituto Superior Técnico, Universidade de Lisboa,\\
  Av. Rovisco Pais 1049-001, Lisboa, Portugal}

\emailAdd{pedromanuelpereira@tecnico.ulisboa.pt}

\abstract{This work analyses a class of extensions of the Standard Model through the lens of a novel exact parameterisation. This parameterisation is specially suitable to the analysis of extensions of the Standard Model with non-unitary mixing matrices, like models with vector-like fermions or right-handed neutrinos. The usefulness of this parameterisation is motivated with two example models: A) Standard Model with the addition of $n_u$ up and $n_d$ down singlet vector-like quarks and B) Standard Model with the addition of $n_R$ right-handed neutrinos. }

\FullConference{%
  Corfu Summer Institute 2021 "School and Workshops on Elementary Particle Physics and Gravity"\\
  29 August - 9 October 2021\\
  Corfu, Greece
}

\begin{document}
\maketitle

\section{Introduction}
This paper is based on refs. \cite{Agostinho:2017wfs,Branco:2020yvs,Branco:2021vhs} as well as work in progress done in collaboration with J.M. Alves, G.C. Branco, A.L. Cherchiglia, C.C. Nishi,  J.T. Penedo,  M.N. Rebelo, and  J.I. Silva-Marcos, to whom I am deeply grateful. \\
Studying a class of extensions of the Standard Model with non-unitary mixing matrices can easily become a very demanding task when the region of interest of the parameter space is very close to regions where common approximations, widely used in the literature, fail. From this problem originated the need to develop an exact parameterisation, that would work in all cases.\\
For instance, the classic neutrino seesaw mechanism papers \cite{Yanagida:1979as,Minkowski:1977sc} work under the assumption that the deviations from unitarity of the leptonic mixing matrix are negligible, which imply that the heavy neutrino masses are around the GUT scale, in order to have order one Yukawa couplings in the neutrino Dirac mass matrix. Furthermore, the famous Casas-Ibarra parameterisation \cite{Casas:2001sr} is also an approximation, if used naively, as explained in section 2.2 of ref. \cite{Branco:2020yvs}. 
Therefore, if one wants to study a model where some of the heavy neutrinos have masses around the GeV scale or below, the aforementioned assumptions start to crumble. Recently, a parameterisation where the mixing matrix is parameterised via a power series was introduced \cite{Fernandez-Martinez:2016lgt}. The small parameter of the power series is the deviations from unitarity of the mixing matrix and a first order truncation on this parameter yields an accurate result for models with GeV neutrinos. However, if one wants to study a model where the spectra of heavy neutrinos includes eV, keV or MeV masses, none of the parameterisations available in the literature was suitable for it. \\
In the case of vector-like quarks, many works in the literature use the assumption that vector-like quarks only couple to the third generation of SM quarks \cite{Aguilar-Saavedra:2013qpa}. This does not need to be the case, as experimental evidence signals that if vector-like quarks are the explanation for the CKM unitarity problem, then the vector like quark might couple more to the 1st generation, as discussed in ref. \cite{Branco:2021vhs}. \\
Hence, the following exact parameterisation is suitable when one wants to study a region of the parameter space where common approximations fail or when one  wants to perform a general scan of the parameter space, without biases.

\section{Mass Matrices and their Diagonalisation}
The usefulness of the parameterisation will be motivated with two example models:

\begin{enumerate}[label=(\Alph*)]
\item SM with the addition of $n_u$ up and $n_d$ down singlet vector-like quarks
\item SM with the addition of $n_R$ right-handed neutrinos.
\end{enumerate}

The procedure to derive the parameterisation will be performed in the next sections. Keep in mind that \textbf{no approximations} will be performed in any step, the results are exact. The first step is to define the mass matrices and then diagonalise them. Considering the mass Lagrangian of \textbf{(A)}

\begin{equation}
\mathcal{L}_\text{m}^q \,=\, 
- \begin{pmatrix}
    \overline{d}^0_L &  \overline{D}^0_L
  \end{pmatrix}
  \,\mathcal{M}_d\,
  \begin{pmatrix}
    d^0_{R} \\[1mm] D^0_{R} 
  \end{pmatrix}
- \begin{pmatrix}
    \overline{u}^0_L &  \overline{U}^0_L
  \end{pmatrix}
  \,\mathcal{M}_u\,
  \begin{pmatrix}
    u^0_{R} \\[1mm] U^0_{R} 
  \end{pmatrix}
+ \text{h.c.}\,,
\end{equation}

and of \textbf{(B)}, in the weak basis (WB) where charged leptons are diagonal,

\begin{equation}
\begin{aligned}
\mathcal{L}_{\text{m}}^\nu \,&=\,
-\left[\frac{1}{2}n_{L}^T C^{*} \cM^* n_{L}
+\overline{l_{L}} d_{l}l_{R}\right]+\text{h.c.}\,, 
\end{aligned}
\end{equation}
where $n_L^{T} = (\nu_L^{T}\,\,\, \overline{\nu_R}\, C^T)$,
$n_L = (\nu_L\,, \,\,C\, \overline{\nu_R}^T)$ is a column vector and $C$ is the charge conjugation matrix. The up or down ($q=u,d$) quark $(3+n_q)\times (3+n_q)$ mass matrix is given by
           \begin{align}
\renewcommand{\arraystretch}{1}
\setlength{\extrarowheight}{6pt}
\mathcal{M}_q \,=\,
\left(\begin{array}{c;{2pt/2pt}c}
\,m_q\,\, & \,\om_q\,
\\[1.5mm] \hdashline[2pt/2pt]
\!{\,\,\,\oM_q\,\,\,}\! &
\!{\,\,\,M_q\,\,\,}\!\\[1mm]
\end{array}\right) \,, 
\end{align}

and the $(3+n_R)\times (3+n_R)$ neutrino mass matrix is given by

\begin{align}
\renewcommand{\arraystretch}{1}
\setlength{\extrarowheight}{6pt}
\mathcal{M} \,=\,
\left(\begin{array}{c;{2pt/2pt}c}
\,\textbf{0}\,\, & \,m\,
\\[1.5mm] \hdashline[2pt/2pt]
{\,\,\,m^T\,\,\,}\! &
{\,\,\,M\,\,\,}\!\\[1mm]
\end{array}\right) \,, 
\end{align}
the diagonalisation equations are
\begin{equation}
 {\mathcal{V}^q_L}^\dagger
\,\,\mathcal{M}_q\,\,
\mathcal{V}_R^q
\,=\, \mathcal{D}_q\, \,~~,~~  \cV^T \cM^* \cV = \cD ~.
\label{eq:diag}
\end{equation}
The matrices that diagonalise the mass matrices, $\mathcal{V}$, are $(3+n)\times (3+n)$ \footnote{$n=n_q$ for quarks and $n=n_R$ for neutrinos}  unitary matrices and can be decomposed into two other matrices $A$ and $B$

           \begin{equation}
\cV^q_\chi \,=\,
\setlength{\extrarowheight}{1.2pt}
\left(\begin{array}{c}
{ }\\[-4mm]
\qquad A^q_\chi \qquad 
\\[2mm] \hdashline[2pt/2pt]
{ }\\[-4mm]
\qquad B^q_\chi \qquad 
\\[2mm]
\end{array}\right) 
\,,
\label{eq:vquarks}
\end{equation}

 \begin{equation}
\cV \,=\,
\setlength{\extrarowheight}{1.2pt}
    \left(\begin{array}{c}
     { }\\[-4mm]
      \qquad A \qquad 
      \\[2mm] \hdashline[2pt/2pt]
       { }\\[-4mm]
      \qquad B\qquad 
      \\[2mm]
    \end{array}\right) 
\,,
\label{eq:vneutrinos}
\end{equation}
where $\chi=L,R$, $A$ is a $3 \times (3+n)$ and $B$ a $n \times (3+n) $ matrix. $A$ and $B$ are non-unitary matrices, in general.
After this decomposition into $A$ and $B$, one can rewrite the equations in eq. \eqref{eq:diag} as
\begin{equation}
        \begin{split}
    &m_q   = A^q_{L} \,\mathcal{D}_q\, {A_{R}^q}^\dagger\,, \quad \\
 & \om_q = A^q_{L} \,\mathcal{D}_q\, {B_{R}^q}^\dagger\,, \quad \\
  & \oM_q = B^q_{L} \,\mathcal{D}_q\, {A_{R}^q}^\dagger\,, \quad \\
  & M_q   = B^q_{L} \,\mathcal{D}_q\, {B_{R}^q}^\dagger\,,
   \end{split}
   \label{eq:mquarks}
\end{equation}
for quarks, and
 
 \begin{equation}
 \begin{split}
&\bzero = A \,\cD\, A^T\,, \quad\\
&m = A \,\cD\, B^T\,, \quad \\
  &M   = B \,\cD\, B^T\,,
 \end{split}
 \label{eq:mneutrinos}
\end{equation}
for neutrinos.\\
The last thing to do is a proper parameterisation of these $A$ and $B$ matrices. Before proceeding, the next section contains, for completeness, the charged currents and neutral interactions Lagrangians for quarks and neutrinos, in the physical basis. Naturally, these will only depend on the $A$ part of the unitary matrices $\mathcal{V}$, as this contains all the physical parameters of the theory, as it will become clear in the \textbf{Parameterisation} section.

\section{Interactions}

The charged current Lagrangian for the quarks is given by
\begin{equation}
   \mathcal{L}_W^q=-\frac{g}{\sqrt{2}}
 \begin{pmatrix}
\overline{u}_L & \overline{U}_L 
\end{pmatrix}\,
V\,    \gamma^\mu
\begin{pmatrix}
d_L \\[2mm] D_L 
\end{pmatrix} 
    W_\mu^+
    \,+\,\text{h.c.} ~,
 \end{equation}
where
       \begin{equation}
           V=A_L^{u^\dag} A_L^d ~,
       \end{equation}
and for the neutrinos
 \begin{equation}
   \mathcal{L}_W^\nu=-\frac{g}{\sqrt{2}}
\overline{l}_L \,
V\,    \gamma^\mu
\begin{pmatrix}
n_L \\[2mm] N_L 
\end{pmatrix} 
    W_\mu^+
    \,+\,\text{h.c.} ~,
 \end{equation}
 where
\begin{equation}
 V=A  ~.
\end{equation}
The part of the Lagrangian that contains Z interactions for the quarks is given by

\begin{equation}
\begin{split}
    \mathcal{L}_Z^q&=-\frac{g}{2 \cos{\theta_W}}  Z_\mu \left[
 \begin{pmatrix}\overline{u}_L & \overline{U}_L \end{pmatrix}
\,F^u\, \gamma^\mu
\begin{pmatrix} u_L \\[2mm] U_L \end{pmatrix}
-
 \begin{pmatrix}\overline{d}_L & \overline{D}_L \end{pmatrix}
\,F^d\, \gamma^\mu
\begin{pmatrix} d_L \\[2mm] D_L \end{pmatrix}
- 2 \sin^2{\theta_W} J^{\mu~q}_\text{em} \right] \,,
\end{split}
\end{equation}
where 

\begin{equation}
\begin{aligned}
J^{\mu~q}_\text{em} &=\, 
\frac{2}{3}  
\Big( \overline{u_{i}} \gamma^\mu  {u_{i}}
+ \overline{U_{r}} \gamma^\mu {U_{r}}  \Big)  
- \frac{1}{3} 
\Big(\overline{d_{i}} \gamma^\mu  {d_{i}}
+ \overline{D_{s}} \gamma^\mu {D_{s}} \Big)
\,,
\end{aligned}
\end{equation}
and $\psi \equiv \psi_L + \psi_R$ for $\psi\in\{u_i,d_i,U_r,D_s\}$, the index $i$ runs from 1 to 3 as in the SM, while the indices $r,s$ run from 1 to $n_{u,d}$.
While the part that contains Higgs interactions is given by
\begin{equation}
 \mathcal{L}_H^q \,=\,
 -\frac{h}{v} \bigg[\begin{pmatrix} \overline{u}_L & \overline{U}_L \end{pmatrix}\,
    F^u\, \mathcal{D}_u\begin{pmatrix} u_R \\[2mm] U_R \end{pmatrix}  + \begin{pmatrix} \overline{d}_L & \overline{D}_L \end{pmatrix}\,
    F^d\, \mathcal{D}_d\begin{pmatrix} d_R \\[2mm] D_R \end{pmatrix}  \bigg]+\,\text{h.c.}
    \,~,
\end{equation}
where
       \begin{equation}
           F^q=A_{L}^{q^\dag} A_{L}^{q} ~.
       \end{equation}
Note that $F^q$ is present in interactions with the Higgs and the Z boson and is the matrix that controls Flavour Changing Neutral Couplings (FCNCs).
For neutrinos,
\begin{equation}
\begin{split}
    \mathcal{L}_Z^\nu&=- \frac{g}{2\cos{\theta_W}}  Z_\mu [ \left( 
\begin{array}{cc}
\overline{n_L} &  \overline{N_L}
\end{array}%
\right) ~F~  \gamma^\mu\left( 
\begin{array}{cc}
n_L  \\ 
N_L%
\end{array}%
\right)] +\,\text{h.c.} ~,\\
\end{split}
\end{equation}

\begin{equation}
 \mathcal{L}_H^\nu \,=\,
 -\frac{h}{v} \bigg[\begin{pmatrix} \overline{n}_L & \overline{N}_L \end{pmatrix}\,
    F~\cD\, \begin{pmatrix} n_L^c \\[2mm] N_L^c \end{pmatrix} 
 \, \bigg] 
 \,+\,\text{h.c.}
    \,,
\end{equation}

where
   \begin{equation}
        F=A^\dag A ~.
   \end{equation}

\section{Parameterisation}
In ref. \cite{Agostinho:2017wfs}, where the neutrino case was studied, it was shown that a unitary matrix $\mathcal{V}$ can be parameterised as

   \begin{align}
\renewcommand{\arraystretch}{1.2}
\mathcal{V}\,=\,
\renewcommand{\arraystretch}{1}
\setlength{\extrarowheight}{6pt}
    \left(\begin{array}{c;{2pt/2pt}c}
      \,\,\,K\,\,\,\,\, & \,\,\,K\,X^\dag\,
      \\[1.5mm] \hdashline[2pt/2pt]
     {- \oK \,{X}\,\, }\! &
{\quad \,\,\oK\quad\,\,}\!\\[1mm]
    \end{array}\right) ~,
\\[-4mm] \nonumber
\end{align}
for a non-singular $3 \times 3$  general complex matrix, $K$, a non-singular $n_R \times n_R$  general complex matrix ~$\oK$ and a $n_R \times 3$ matrix, $X$, that will be defined next. From eq. \eqref{eq:vneutrinos} one can identify
\begin{equation}
    A = ( K~~KX^\dag ) ~,~ B = ( -\oK X~~\oK ) ~,
    \label{eq:ABdef}
\end{equation}
where the unitarity of $\mathcal{V}$ yields
\begin{align}
\mathcal{V} \mathcal{V}^\dag= \begin{pmatrix}
 \,A\, \\[2mm] 
B
\end{pmatrix} \begin{pmatrix}
\,{A}^\dagger & {B}^\dagger\,
\end{pmatrix} 
\,&=\,
\begin{pmatrix}
\,A\,{A}^\dagger & A\,{B}^\dagger\,\\[2mm] 
\,B\,{A}^\dagger & B\,{B}^\dagger\,
\end{pmatrix} 
\,=\,
\begin{pmatrix}
\,\id_{3}\, & \,0\, \\[2mm] 
\,0\, & \,\id_{n_R}\,
\end{pmatrix}\,, 
 \label{eq:unitAB02n}
\\[2mm]
\mathcal{V}^\dag \mathcal{V}= \begin{pmatrix}
\,{A}^\dagger & {B}^\dagger\,
\end{pmatrix} \begin{pmatrix}
\,A\, \\[2mm] 
B
\end{pmatrix}
\,&=\,
\,
{A}^\dagger\,A+ 
{B}^\dagger\,B
\,
\,=\,
\,
\id_{3+n_R}
\,.
 \label{eq:unitAB2n}
\end{align}

Using eq. \eqref{eq:ABdef} and the unitarity relations in eqs. \eqref{eq:unitAB02n} and \eqref{eq:unitAB2n}, eq. \eqref{eq:mneutrinos} becomes

 \begin{equation}
 \begin{aligned}
& \bzero = d + X^\dag D X^* ~,\\
&m \,=\, K\, X^\dagger D \left( \oK ^{-1}\right)^* ~,\\
  &M \,=\, \oK \, (D+X\,d\,X^T)\,\oK ^T\,.
 \end{aligned}
 \label{eq:mneutrino}
\end{equation}
The last two equations are exact equations for the $3 \times n_R$ Dirac mass matrix, $m$, and the $n_R \times n_R$ Majorana mass matrix, $M$.
The equation involving the null matrix has the solution
\begin{align}
X\,=\pm\,i\,\sqrt{D^{-1}}\,O_{c}\,\sqrt{d}\,,
\end{align}%
where $O_c$ is an orthogonal complex matrix and $d$ ($D$) is a diagonal matrix with the masses of the light (heavy) neutrinos in the diagonal. This unique solution defines the matrix $X$. From the unitarity relations of $\mathcal{V}$ one can also obtain the following definitions
 \begin{equation}
 \begin{split}
&K=U_{K}(\id_3+X^\dag X)^{-1/2} ~,\\
&\oK=U_{\oK}(\id_{n_R}+X X^\dag)^{-1/2}~,\\
&K_{PMNS}= K~,
 \end{split}
\end{equation}
where $U_K$ and $U_{\oK}$ are unitary matrices, since the unitarity relations only define $K$
 and $\oK$ up to a unitary matrix on the left. $K$ will play the role of the PMNS mixing matrix and is only unitary when $X \xrightarrow[]{} \bzero$.\\

 For the quarks, one can perform the same steps and write

           \begin{align}
\renewcommand{\arraystretch}{1.2}
\mathcal{V}_\chi^q \,=\,
\renewcommand{\arraystretch}{1}
\setlength{\extrarowheight}{6pt}
\left(\begin{array}{c;{2pt/2pt}c}
\,\,\,K_\chi^q\,\,\,\,\, & \,\,\,K_\chi^q\,{X_\chi^q}^\dag\,
\\[1.5mm] \hdashline[2pt/2pt]
{- \oK_{\!\chi}^{\!q}\,X_\chi^q\,\, }\! &
{\quad \,\,\oK_{\!\chi}^{\!q}\quad\,\,}\!\\[1mm]
\end{array}\right) ~.
\end{align}
 for a non-singular $3 \times 3$  general complex matrix, $K^q_\chi$, a non-singular $n_q \times n_q$  general complex matrix ~$\oK^q_\chi$ and a $n_q \times 3$ matrix, $X^q_\chi$. Again, from eq. \eqref{eq:vquarks} one obtains
\begin{equation}
    A^q_\chi = ( K^q_\chi~~K^q_\chi {X^q_\chi}^\dag ) ~,~ B^q_\chi = ( -\oK^q_\chi X^q_\chi~~~\oK^q_\chi ) ~,
    \label{eq:ABdefquark}
\end{equation}
where the unitarity of $\mathcal{V}^q_\chi$ yields
\begin{align}
\mathcal{V}^q_\chi {\mathcal{V}^q_\chi}^\dag = \begin{pmatrix}
\,A^q_\chi\, \\[2mm] 
B^q_\chi
\end{pmatrix} \begin{pmatrix}
\,{A^q_\chi}^\dagger & {B^q_\chi}^\dagger\,
\end{pmatrix} 
\,&=\,
\begin{pmatrix}
\,A^q_\chi\,{A^q_\chi}^\dagger & A^q_\chi\,{B^q_\chi}^\dagger\,\\[2mm] 
\,B^q_\chi\,{A^q_\chi}^\dagger & B^q_\chi\,{B^q_\chi}^\dagger\,
\end{pmatrix} 
\,=\,
\begin{pmatrix}
\,\id_{3}\, & \,0\, \\[2mm] 
\,0\, & \,\id_{n_q}\,
\end{pmatrix}\,, 
 \label{eq:unitAB02}
\\[2mm]
 {\mathcal{V}^q_\chi}^\dag \mathcal{V}^q_\chi=\begin{pmatrix}
\,{A^q_\chi}^\dagger & {B^q_\chi}^\dagger\,
\end{pmatrix} \begin{pmatrix}
\,A^q_\chi\, \\[2mm] 
B^q_\chi
\end{pmatrix}
\,&=\,
\,
{A^q_\chi}^\dagger\,A^q_\chi+ 
{B^q_\chi}^\dagger\,B^q_\chi
\,
\,=\,
\,
\id_{3+n_q}
\,,
 \label{eq:unitAB2}
\end{align}
for each $q=u,d$ and $\chi = L,R$.

Using eq. \eqref{eq:ABdefquark} and the unitarity relations in eqs. \eqref{eq:unitAB02} and \eqref{eq:unitAB2}, eq. \eqref{eq:mquarks} becomes

\begin{equation}
\begin{split}
&m_q \,=\,K_L^q \left( d_q + {X_L^q}^\dag\,D_q\,X_R^q \right ) {K_R^q}^\dagger, \\
 & \om_q \,=\, K_L^q \left( {X_L^q}^\dag\,D_q - d_q\,{X_R^q}^\dag \right ) {\oK_R^q}^\dagger,\\
  & \oM_q \,=\,\oK_L^q \left( D_q\,X_R^q - X_L^q\,d_q \right ) {K_R^q}^\dagger , \\
  & M_q \,=\,\oK_L^q \left( D_q + X_L^q\,d_q\,{X_R^q}^\dag \right ) {\oK_R^q}^\dagger,
\end{split}
\label{eq:mquark}
\end{equation}
where $m_q$ is the $3 \times 3$ Dirac mass matrix for the quarks and the entries of the $3 \times n_q$ matrix, $\om_q$, are proportional to the Higgs vacuum expectation value, as well. The $n_q \times 3$ mass matrix, $\oM_q$, and the $n_q \times n_q$ mass matrix, $M_q$, are bare mass terms involving only singlet quark fields.\\
It is always possible to go to a WB where $\om_q$ is $\textbf{0}$ \footnote{This is always possible for $\oM_q$, as well.}. In that WB, one can proceed like in the neutrino case to obtain a formula for $X_\chi^q$,
\begin{eqnarray}
\begin{aligned}
&X_L^q =  \sqrt{D_q^{-1}}    P^q   \sqrt{d_q} ~,\\
&X_R^q = \sqrt{D_q} P^q \sqrt{d_q^{-1}} ~,
\end{aligned}
\end{eqnarray}
where $P^q$ is a general complex matrix and $d_q$ ($D_q$) is a diagonal matrix with the masses of the light (heavy) quarks in the diagonal. The following definitions can also be obtained from the unitarity relations in eqs. \eqref{eq:unitAB02} and \eqref{eq:unitAB2},
\begin{equation}
\begin{split}
&K^q_\chi=U^{q}_{K\chi}(\id_3+{X_\chi^q}^\dag X_\chi^q)^{-1/2}\,,\\
&\oK^q_\chi=U^{q}_{\oK\chi}
(\id_{n_q}+ X_\chi^q {X_\chi^q}^\dag)^{-1/2}~,\\
&K_{CKM}= {K_L^u}^\dag K_L^d ~,
\end{split}
\end{equation}
where $U^{q}_{K \chi}$ and $U^{q}_{\oK \chi}$ are unitary matrices, since the unitarity relations only define $K$
 and $\oK$ up to a unitary matrix on the left. The combination ${K_L^u}^\dag K_L^d$ will play the role of the CKM mixing matrix and is only unitary when $X_L^u, X_L^d \xrightarrow[]{} \bzero$.\\

The matrices relevant for FCNCs, involving the $Z$ and the Higgs, as defined in the section \textbf{Interactions}, are given by

 \begin{equation}
F^q= \begin{pmatrix}
(\id_3+{X_L^q}^\dag X_L^q)^{-1} &(\id_3+{X_L^q}^\dag {X_L^q})^{-1}{X_L^q}^\dag\\ 
{X_L^q} (\id_3+{X_L^q}^\dag {X_L^q})^{-1} &{X_L^q} (\id_3+{X_L^q}^\dag {X_L^q})^{-1} {X_L^q}^\dag\\ 
\end{pmatrix} ~,
\end{equation}
for quarks, and
 \begin{equation}
F= \begin{pmatrix}
(\id_3+X^\dag X)^{-1} &(\id_3+X^\dag X)^{-1}X^\dag\\ 
X (\id_3+X^\dag X)^{-1} &X (\id_3+X^\dag X)^{-1} X^\dag\\ 
\end{pmatrix} ~,
\end{equation}
for neutrinos.\\
Finally, the exact formula for the $(3+n) \times (3+n)$ unitary matrix assumes the form
  \begin{equation}
  \mathcal{V}^q_\chi =       \begin{pmatrix}

  U^{q}_{K\chi}(\id_3+{X_\chi^q}^\dag X_\chi^q)^{-1/2} &U^{q}_{K\chi}(\id_3+{X_\chi^q}^\dag X_\chi^q)^{-1/2}X_\chi^{q\dag}\\ 
- U^{q}_{\oK\chi}
(\id_{n_q}+ X_\chi^q {X_\chi^q}^\dag)^{-1/2}X^q_\chi &U^{q}_{\oK\chi}
(\id_{n_q}+ X_\chi^q {X_\chi^q}^\dag)^{-1/2}\\
\end{pmatrix} ~,
\end{equation}  
for quarks, and
 \begin{equation}
  \mathcal{V} =       \begin{pmatrix}
U_{K}(\id_3+X^\dag X)^{-1/2} &U_{K}(\id_3+X^\dag X)^{-1/2}X^\dag\\ 
- U_{\oK}(\id_{n_R}+X X^\dag)^{-1/2}X &U_{\oK}(\id_{n_R}+X X^\dag)^{-1/2}\\
\end{pmatrix} ~,
\end{equation}  
for neutrinos.\\
Note that parameterisations with a similar structure, used in the leptonic sector, existed in the literature prior to this work \cite{Korner:1992zk}\cite{Grimus:2000vj}, but are either approximations or a special case of this one.

\section{Procedure and Usefulness}
Now that all the necessary equations were derived, one needs to obtain observables out of them. One procedure is the following
\begin{itemize}
    \item Start with $d$, $D$, $U_K$ and $O_c$/$P^q$,
    \item Calculate $X$,
    \item Calculate mass matrices, charged currents matrix $V$ and the FCNCs matrix $F$.
\end{itemize}
With these, one should be able to obtain any observable, exact at tree level. Another valid approach is to proceed in the reverse order, but starting from the mass matrices would be inherently more difficult due to the redundant extra parameters.\\
It is important to emphasize that all physical parameters are contained in $d$, $D$, $U_K$ and $O_c$/$P^q$. For the neutrino case, a counting for $n_R=3$ was performed in table 1 of ref. \cite{Branco:2020yvs}.\\
Performing the depicted prescription yields results at tree level that are exact. However, there are some caveats.\\
In the case of neutrinos, one should mind radiative corrections on the light neutrino masses, as stated in section 4 of ref. \cite{Branco:2020yvs}. The upshot from that discussion is that one needs some kind of lepton-number like softly broken symmetry to protect these radiative corrections of becoming too large, when the mass of the heavy neutrinos is below the GeV scale. This is known in the literature as “symmetry protected seesaw models” \cite{Antusch:2015mia}.\\
A concern that exists in every model is perturbativity. This is easy to understand when one analyses the Dirac mass matrix, $m$, equation in \eqref{eq:mneutrino} or \eqref{eq:mquark}: If the Yukawa couplings\footnote{$m/v$, where $v$ is the Higgs vacuum expectation value.} are $O(1)$ then $X^\dag D$ must also be $O(1)$, which implies that deviations from unitarity need to  decrease when the masses of the heavy neutrinos increase, or the inverse. This fact was used to set an upper bound on the mass of the hypothetical up vector-like quark singlet introduced in ref. \cite{Branco:2021vhs} to solve the CKM unitarity problem.\\
The perturbativity issue was discussed in some detail, for quarks, on section 2.5 of ref. \cite{Branco:2021vhs} and, for neutrinos, on section 2.3 of ref. \cite{Branco:2020yvs}.

\section{Conclusions}
In this article, two models with non-unitary mixing matrices were analysed through the lens of an exact parameterisation that proved to be extremely useful. The usefulness of this parameterisation resides in the exact formulas at tree level, which are easy to implement numerically. Moreover, deviations from unitarity of the mixing matrix are controlled by the matrix $X$ and no approximations regarding them or the mass scale of heavy fields is needed to obtain exact results. Furthermore, it can be used in any model with non-unitary mixing matrices: Inverse Seesaw, Linear Seesaw, type-II and type-III seesaw or models with vector like fermions and scalars.\\
More details regarding this analysis can be found in the following works \cite{Agostinho:2017wfs,Branco:2020yvs,Branco:2021vhs}.\\

\section*{Acknowledgements}
This work was partially supported by Fundação para a Ciência e a Tecnologia (FCT, Portugal) through the
projects CFTP-FCT Unit 777 (UIDB/00777/2020 and UIDP/00777/2020), PTDC/FISPAR/29436/2017, CERN/FIS-PAR/0004/2019 and CERN/FIS-PAR/0008/2019, which are partially funded through POCTI (FEDER), COMPETE, QREN and EU. P.M.F.P. acknowledges support from FCT through the PhD grant SFRH/BD/145399/2019


\begin{thebibliography}{99}

\bibitem{Branco:2020yvs}
G.~C.~Branco, J.~T.~Penedo, P.~M.~F.~Pereira, M.~N.~Rebelo and J.~I.~Silva-Marcos,
``Type-I Seesaw with eV-Scale Neutrinos,''
JHEP \textbf{07} (2020), 164
doi:10.1007/JHEP07(2020)164
[arXiv:1912.05875 [hep-ph]].

\bibitem{Branco:2021vhs}
G.~C.~Branco, J.~T.~Penedo, P.~M.~F.~Pereira, M.~N.~Rebelo and J.~I.~Silva-Marcos,
``Addressing the CKM unitarity problem with a vector-like up quark,''
JHEP \textbf{07} (2021), 099
doi:10.1007/JHEP07(2021)099
[arXiv:2103.13409 [hep-ph]].

\bibitem{Agostinho:2017wfs}
N.~R.~Agostinho, G.~C.~Branco, P.~M.~F.~Pereira, M.~N.~Rebelo and J.~I.~Silva-Marcos,
``Can one have significant deviations from leptonic 3 $\times $ 3 unitarity in the framework of type I seesaw mechanism?,''
Eur. Phys. J. C \textbf{78} (2018) no.11, 895
doi:10.1140/epjc/s10052-018-6347-2
[arXiv:1711.06229 [hep-ph]].


\bibitem{Yanagida:1979as}
T.~Yanagida,
``Horizontal gauge symmetry and masses of neutrinos,''
Conf. Proc. C \textbf{7902131} (1979), 95-99
KEK-79-18-95.

\bibitem{Minkowski:1977sc}
P.~Minkowski,
``$\mu \to e\gamma$ at a Rate of One Out of $10^{9}$ Muon Decays?,''
Phys. Lett. B \textbf{67} (1977), 421-428
doi:10.1016/0370-2693(77)90435-X

\bibitem{Casas:2001sr}
J.~A.~Casas and A.~Ibarra,
``Oscillating neutrinos and $\mu \to e, \gamma$,''
Nucl. Phys. B \textbf{618} (2001), 171-204
doi:10.1016/S0550-3213(01)00475-8
[arXiv:hep-ph/0103065 [hep-ph]].

\bibitem{Fernandez-Martinez:2016lgt}
E.~Fernandez-Martinez, J.~Hernandez-Garcia and J.~Lopez-Pavon,
``Global constraints on heavy neutrino mixing,''
JHEP \textbf{08} (2016), 033
doi:10.1007/JHEP08(2016)033
[arXiv:1605.08774 [hep-ph]].

\bibitem{Aguilar-Saavedra:2013qpa}
J.~A.~Aguilar-Saavedra, R.~Benbrik, S.~Heinemeyer and M.~P\'erez-Victoria,
``Handbook of vectorlike quarks: Mixing and single production,''
Phys. Rev. D \textbf{88} (2013) no.9, 094010
doi:10.1103/PhysRevD.88.094010
[arXiv:1306.0572 [hep-ph]].

\bibitem{Korner:1992zk}
J.~G.~Korner, A.~Pilaftsis and K.~Schilcher,
``Leptonic CP asymmetries in flavor changing H0 decays,''
Phys. Rev. D \textbf{47} (1993), 1080-1086
doi:10.1103/PhysRevD.47.1080
[arXiv:hep-ph/9301289 [hep-ph]].

\bibitem{Grimus:2000vj}
W.~Grimus and L.~Lavoura,
``The Seesaw mechanism at arbitrary order: Disentangling the small scale from the large scale,''
JHEP \textbf{11} (2000), 042
doi:10.1088/1126-6708/2000/11/042
[arXiv:hep-ph/0008179 [hep-ph]].

\bibitem{Antusch:2015mia}
S.~Antusch and O.~Fischer,
``Testing sterile neutrino extensions of the Standard Model at future lepton colliders,''
JHEP \textbf{05} (2015), 053
doi:10.1007/JHEP05(2015)053
[arXiv:1502.05915 [hep-ph]].



\end{thebibliography}
\end{document}